\documentstyle[aps,epsfig]{revtex}

\title{Description of quantum dynamics of open systems based on collision-like models}
\author{
M\'ario Ziman$^{1,2}$, Peter \v Stelmachovi\v c$^{1}$, Vladim\'\i r Bu\v zek$^{1,2}$}
\address{
$^1$Research Center for Quantum Information,
  Slovak Academy of Sciences,
  D\'ubravsk\'a cesta 9, 845 11 Bratislava, Slovakia \\
$^2$Faculty of Informatics, Masaryk University, Botanick\'a 68a, 602 00 Brno,
  Czech Republic
}

\begin{document}

\maketitle
\begin{abstract}

Master equations in the Lindblad form describe evolution of open quantum systems that is completely positive
and simultaneously has a semigroup property. We analyze a possibility to derive this type
of master equations from an intrinsically
discrete dynamics that is modelled as a sequence of collisions between a given quantum system
(a qubit) with particles that form the environment.
In order to illustrate our approach we analyze in detail how a process of an exponential decay and
a process of decoherence can be derived from a
collision-like model in which particular collisions are described by SWAP and controlled-NOT interactions, respectively.
\end{abstract}
\section{Introduction}

The central issue in the study of evolution of open systems
is the observed {\it irreversibility} of both classical
as well as quantum dynamics. Open systems interacting
with an environment gradually loose their information
content and decohere, which means that after some time their
states are  to some extent  determined by the initial parameters of the
environment. Such behavior cannot be described by unitary
transformations and has led to an introduction of a phenomenological
dynamical postulate for open systems \cite{lindblad76,davies,alicki} - the {\it semigroup
property} of the time evolution ${\cal E}_t$
\begin{equation}
\label{semigroup}
{\cal E}_{t+s} = {\cal E}_t {\cal E}_s\; ; \hspace{1cm} \forall t , s \geq 0 \; ,
\ \ \ {\rm and}\ \ \ \lim_{t\to 0^+}{\cal E}_t={\cal I} \; ,
\end{equation}
where ${\cal E}_t: {\cal S} ({\cal H}) \rightarrow {\cal S} ({\cal H})$
is a transformation acting on the set of all possible states
${\cal S}({\cal H})$ of a given quantum system corresponding
to the Hilbert space ${\cal H}$. In this case it is possible
that pure quantum states evolve into mixtures, and vice versa.

The irreversibility is related to the non-existence
of the inverse evolution ${\cal E}_t^{-1}$.
In particular, the inverse map can exist in the mathematical sense, but
the resulting map does not describe a valid quantum evolution.
The physical maps ${\cal E}_t$ must satisfy
several constraints \cite{Nielsen2000}: They have to be linear, trace-preserving and
they have to be completely positive for all $t$.

The state of the system $\varrho_t$ is obtained by the application of the
map on the initial state $\varrho$, i.e. $\varrho_t={\cal E}_t[\varrho]$.
Equations of motion that describe this type of dynamics has the
so-called {\it Lindblad form} \cite{lindblad76}:
\begin{equation}
\dot{\varrho}_t={\cal G}[\varrho_t]=-i[H,\varrho_t]
+\frac{1}{2}\sum_{\alpha,\beta}c_{\alpha\beta}
\left(
[\Lambda_\alpha,\varrho_t\Lambda_\beta]+
[\Lambda_\alpha\varrho_t,\Lambda_\beta]
\right)\; ,
\end{equation}
where $\Lambda_\alpha=\Lambda_\alpha^\dagger$, ${\rm Tr}\Lambda_\alpha=0$,
${\rm Tr}\Lambda_\alpha\Lambda_\beta=\delta_{\alpha\beta}$
for $\alpha,\beta=1,\dots,d^2-1$ and the coefficients $c_{\alpha\beta}$
form a positive matrix. The Hamiltonian $H$ is traceless, i.e.
$H=\sum_\alpha h_\alpha\Lambda_\alpha$. This differential equation
is supplemented by the initial condition $\varrho_{t=0}=\varrho$.

Differentiating the time evolution $\varrho_t={\cal E}_t[\varrho]$
we derive a formal expression for the generator of the dynamics $\cal G$
\begin{equation}
\dot{\varrho}_t=\dot{\cal E}_t[\varrho]=
\dot{\cal E}_t{\cal E}_t^{-1}[\varrho_t]\ \Rightarrow\
{\cal G}=\dot{\cal E}_t{\cal E}_t^{-1}\; .
\label{generator}
\end{equation}
This expression in a strict mathematical sense has a meaning only when inverse transformations
${\cal E}_t^{-1}$ do exist for any time $t$. For general one-parametric set of
completely positive maps ${\cal E}_t$ the generator can be time-dependent,
however, if these maps possess the semigroup property, then
the generator is independent of time.

\section{Discrete dynamical semigroup}

The derivation of the particular master equation, which drives the
evolution of open system, is usually based on the idea that the open system
is a part of the larger closed system with the underlying unitary dynamics
described by Schr\"odinger equation. However, after tracing out
the environment, the induced set of maps ${\cal E}_t$
essentially never fulfills the semigroup property. In order to obtain
this feature of dynamics various approximations have 
to be applied \cite{alicki,spohn},
for instance the Born-Markov approximation, the weak-coupling limit, the mean-field limit,
etc. Master equations of the Lindblad form
are valid only under specific physical conditions, but still provide
us with a very reasonable approximative picture of the exact dynamics of open quantum systems.

\begin{figure}
\begin{center}
\includegraphics[width=7cm]{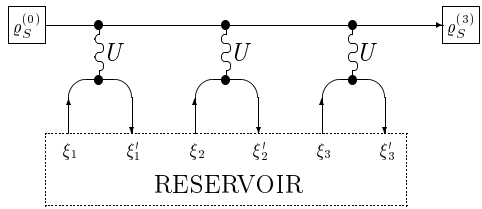}
\caption{A schematic visualization of a sequence of collisions
between the system and particles from the environment. This interactions lead to a dynamics
of the system that is described by a specific sequence of maps ${\cal E}_n$ that in the continuous limit
can be described by a specific master equation.}
\label{col_model}
\end{center}
\end{figure}

In this paper we will present a novel method how to derive
master equations. We will consider that the interaction between
the system and its environment consists of bipartite collisions. For this we assume that
the environment consists of $N$ particles (quantum systems) of the same physical origin as the system
under consideration. In our model we assume that the environment is initially ``prepared'' in a factorized state
$\xi^{\otimes N}$ and the system interacts with each particle
from the environment at most once. Each collision is described by
a unitary transformation $U$, which induces a map ${\cal E}$
(see Fig.\ref{col_model}).
As a result this {\it collision-like} model determines a discrete evolution
described by powers of the map $\cal E$, i.e. ${\cal E}^n$
with $n=1,2,\dots$ corresponding to the $n$th collision. The set
of maps ${\cal E}_n:={\cal E}^n$ with ${\cal E}_0={\cal I}$
obviously fulfills the semigroup property, i.e.
\begin{equation}
{\cal E}_n{\cal E}_m={\cal E}_{n+m} \ \ \ {\rm for\ all}\ \ \
n,m=0,1,2,\dots
\end{equation}
In other words the maps ${\cal E}_n$ form a {\it one-parametric
discrete semigroup}.

The question is whether we may replace the discrete dynamics
${\cal E}_n$ with a continuous one ${\cal E}_t$ such that
${\cal E}_{t_n}={\cal E}_n$ for times $t_n=n\tau$
($\tau$ is some fixed time interval between subsequent collisions),
and the maps ${\cal E}_t$ satisfy the semigroup property. In other words
the task is to interpolate a discrete set of points with a line
in the abstract space of quantum maps. Obviously, on the time scale
less than $\tau$ the continuous evolution is different
from the discrete one. However, our interest is to describe the
overall evolution rather than a sequence of isolated  collision.
In this sense, the continuous
dynamics is a good approximation of the discrete one. Our aim is to
describe the intrinsically discrete collision model in the language
of quantum dynamical semigroups, i.e. continuous master equations.

We derive the master equation in the following way.
Firstly, we will express the powers of the map ${\cal E}$ in a specific form. The discrete parameter
$n$ numbering the order of interaction of the system with environment particles
 will be replaced by a continuous parameter
$t=n\tau$, i.e. $n\to t/\tau$. Correspondingly, we have
a continuous set of maps ${\cal E}_t$. The question is,
whether these maps are completely positive, and whether they
form a semigroup. If, moreover, the inverse maps ${\cal E}_t^{-1}$
exist, then we can derive the master equation by using
the expression for the generator in Eq. (\ref{generator}).

In what follows we will present two specific  examples, for which the
derived master equations correspond to an exponential decay, and to a
decoherence process.
It is important  to understand which bipartite interactions
underly these type of processes. We will be interested
only in two-dimensional systems, i.e. qubits.

\section{Qubit formalism}
In what follows  we will use the so-called
{\it left-right formalism}, or real matrix representation of
quantum maps, and real vector representation of quantum states
of a qubit. The state space of a qubit
is a subset of four-dimensional linear space of hermitian operators.
It follows that any density operator can be written
as $\varrho=\frac{1}{2}{\it I}+\vec{r}\cdot\vec{\sigma}$,
where $\vec{\sigma}=(\sigma_x,\sigma_y,\sigma_z)$ are the
{\it Pauli operators}. The positivity of $\varrho$
restricts the choices of vector $\vec{r}$ such that $|\vec{r}|\le 1/2$, i.e.
the states of qubit form a three-dimensional sphere - {\it Bloch sphere}.

The quantum evolution maps ${\cal E}$ as well as the generators ${\cal G}$
of quantum
dynamics ${\cal E}$ then correspond to 4x4 real matrices acting on
the vectors $(1/2,\vec{r})$. In particular, the evolution ${\cal E}$ has
the affine form
$
{\cal E}=\left(
\begin{array}{cc}
1 & \vec{0} \\
\vec{t} & T
\end{array}
\right)
$,
where $T$ is some 3x3 matrix and $\vec{t}$ is the translation vector.
Under the action of ${\cal E}$ the Bloch sphere vectors $\vec{r}$ transform
in the following way $\vec{r}\to\vec{r}^\prime=T\vec{r}+\vec{t}$.
The generator $\cal G$ of the dynamics has very similar form, only
the  first row of the matrix vanishes, i.e.
$
{\cal G}=\left(
\begin{array}{cc}
0 & \vec{0} \\
\vec{g} & G
\end{array}
\right)
$.
The matrix elements are given by a relation
$[{\cal G}]_{jk}=\frac{1}{2}{\rm Tr}(\sigma_j{\cal G}[\sigma_k])$ for
$j,k=0,1,2,3$ and $\sigma_0={\it I}$.

For our purposes it will be useful to know how
to rewrite this matrix form of the generator into the operator
form
\begin{equation}
\label{qubit_gen}
\dot{\varrho}_t=-i[H,\varrho_t]+\frac{1}{2}\sum_{j,k=1}^3
c_{jk}\left(
[\sigma_j,\varrho_t\sigma_k]+[\sigma_j\varrho_t,\sigma_k]
\right)\; ,
\end{equation}
i.e. how to rewrite the coefficients $[{\cal G}]_{jk}$
into the parameters $h_j$ and $c_{jk}$.
The hermitian matrix $c_{jk}$ can be rewritten as $c_{jk} = d_{jk} - i e_{jk}$,
where $d_{jk}=\frac{1}{2}(c_{jk}+c_{kj})$ is the real symmetric
matrix and $e_{jk}=i\frac{1}{2}(c_{jk}-c_{kj})$ is the real
antisymmetric matrix. Using the operator expression of the generator
(\ref{qubit_gen}),
one can easily find the matrix
\begin{equation}
{\cal G}=\left ( \begin{array}{cccc}
0 & 0 & 0 & 0 \\
4 e_{23} & -2 d_{22}- 2d_{33}  & \; \; 2 d_{12}-2h_3 \; \; & 2 d_{13} + 2h_2 \\
4 e_{31} & 2 d_{12} + 2 h_3 & -2d_{11} -2  d_{33}  & 2 d_{23} -2h_1 \\
4 e_{12} & \; \; 2 d_{31}-2h_2 \; \;  & 2 d_{32} + 2h_1 & -2 d_{11} -  2 d_{22}
\end{array} \right )\; .
\end{equation}
The inverse relations then read
\begin{equation}
\begin{array}{ccc}
h_1 =  \frac{[{\cal G}]_{32} -[{\cal G}]_{23}}{4}\; ; &
h_2  =  \frac{[{\cal G}]_{13} -[{\cal G}]_{31}}{4}\; ; &
h_3  =  \frac{[{\cal G}]_{21} -[{\cal G}]_{12}}{4}\; ; \\
e_{23}  =  \frac{[{\cal G}]_{10}}{4}\; ;  &
e_{31}  =  \frac{[{\cal G}]_{20}}{4}\; ;  &
e_{12}  =  \frac{[{\cal G}]_{30}}{4}\; ,
\end{array}
\label{inverse1}
\end{equation}
and
\begin{equation}
\begin{array}{cc}
\begin{array}{rcl}
d_{11}   &=&   \frac{-[{\cal G}]_{22}-[{\cal G}]_{33}+[{\cal G}]_{11}}{4}\; ; \\
d_{22}  &=&  \frac{-[{\cal G}]_{11}-[{\cal G}]_{33}+[{\cal G}]_{22}}{4}\; ; \\
d_{33}  &=&  \frac{-[{\cal G}]_{11}-[{\cal G}]_{22}+[{\cal G}]_{33}}{4}\; ;
\end{array}
&
\begin{array}{rcl}
d_{12}  &=&  \frac{[{\cal G}]_{12}+[{\cal G}]_{21}}{4}\; ; \\
d_{23}  &=&  \frac{[{\cal G}]_{23}+[{\cal G}]_{32}}{4}\; ; \\
d_{13}  &=&  \frac{[{\cal G}]_{13}+[{\cal G}]_{31}}{4}\; .
\end{array}
\end{array}
\label{inverse2}
\end{equation}
It is important to note
that these relations enable us to represent not only dynamical semigroups,
but any time evolution of a qubit in the Lindbland-like form. In this
general case the coefficients $c_{jk}$ will not be constant,
but will explicitly depend on time. In this way we can derive
more general master equations describing dynamics beyond Markovian approximation.

\section{Case study I.: Quantum homogenization}
The {\it quantum homogenization} \cite{Ziman012,Scarani01,Ziman2003} is a process
motivated by the thermodynamical process of
{\it thermalization}. It describes a system-reservoir interaction in which the
initial state of the system $\varrho$ is transformed into the state $\xi$
determined by the state of the reservoir that is composed of $N$ systems
of the same physical origin as the system.
The interaction between the system and the reservoir consists
of bipartite collisions. Each collision is described by some unitary map $U$.
In order to obtain a discrete semigroup describing the
dynamics of the system, we assume that initially the reservoir
is in a factorized state $\xi^{\otimes N}$ and that the system interacts
with each system from the reservoir maximally once (see Fig.\ref{col_model}).

The homogenization approximate the evolution
$\varrho\otimes\xi^{\otimes N}\to\xi^{\otimes (N+1)}$, which
is forbidden by the  {\it no-cloning theorem} (see Ref.~\cite{Buzek1996} and references therein).
Let $\delta$ be the parameter describing the quality of
the homogenization in the following sense. After the homogenization process
is complete, all  systems are described by states, which belong to
a $\delta$-vicinity of the state $\xi$. Moreover, the homogenization
requires the validity of the following relations corresponding to
{\it trivial homogenization}
${\rm Tr}_1(U\xi\otimes\xi U^\dagger)=
{\rm Tr}_2(U\xi\otimes\xi U^\dagger)=\xi$.
If we assume that the homogenization
is independent on an initial state of the system qubit ($\varrho$)
and as well as on
initial states of reservoir qubits ($\xi$)  then
for qubits $U$ must posses the form of the {\it partial swap} operation
(for more details see Ref.~\cite{Ziman012})
\begin{equation}
P_\eta = \cos\eta I+i\sin\eta S
\end{equation}
where $S$ is the {\it swap operation} defined by the relation
$S|\psi\rangle\otimes|\phi\rangle=|\phi\rangle\otimes|\psi\rangle$
for all $|\psi\rangle,|\phi\rangle$.
In what follows we will use the notation $c=\cos\eta$ and $s=\sin\eta$.

Let us define the Bloch-sphere vectors for density operators, i.e. $\varrho\leftrightarrow\vec{r}$
and $\xi\leftrightarrow\vec{t}$. In terms of these vectors
the partial-swap induces the map
$\vec{r}\to\vec{r}^\prime=c^2\vec{r}+s^2\vec{t}-2cs\vec{t}\times\vec{r}$, i.e.
the superoperator ${\cal E}_\xi$ is represented by the matrix
\begin{equation}
\label{homosuperoperator}
{\cal E}_\xi=\left(\begin{array}{cccc}
1 & 0 & 0 & 0\\
2s^2 t_x & c^2 & 2cst_z & -2cst_y\\
2s^2 t_y & -2cst_z & c^2 & 2cst_x\\
2s^2 t_z & 2cst_y & -2cst_x & c^2\\
\end{array}\right).
\end{equation}

Now we turn our attention on the derivation of the master equation that corresponds to
a dynamics induced by a sequence of partial-swap interactions.
By a suitable (unitary) substitution of the operator basis
$I,\sigma_x,\sigma_y,\sigma_z\to{\it I},S_x,S_y,S_z$,
such that $S_k=U\sigma_k U^\dagger$ ($UU^\dagger=U^\dagger U={\it I}$),
the matrix ${\cal E}_\xi$ can be rewritten into the form
\begin{equation}
{\cal E}_\xi=\left(\begin{array}{ccc}
1 & 0 \ \ \ 0 & 0\\
\begin{array}{c}0\\ 0 \end{array}
& cA &
\begin{array}{c}0\\ 0 \end{array} \\
2s^2 w & 0 \ \ \ 0 & c^2\\
\end{array}\right) \ \ {\rm with}\ \ \
A=\left(
\begin{array}{cc}
c & 2sw \\
-2sw & c
\end{array}
\right)\, ,
\end{equation}
where $w$ is determined by
the initial state of the reservoir, i.e. $\xi=\frac{1}{2}{\it I}+w S_z$.
This change of the operator basis corresponds to a different choice of the
$x,y,z$ axes in the Bloch sphere.
In what follows we will describe the process in this new
basis, in which the powers of ${\cal E}_\xi$ can be easily find
\begin{equation}
{\cal E}_\xi^n=\left(\begin{array}{ccc}
1 & 0 \ \ \ 0 & 0\\
\begin{array}{c}0\\ 0 \end{array}
& c^n A^n &
\begin{array}{c}0\\ 0 \end{array} \\
2w(1-c^{2n}) & 0 \ \ \ 0 & c^{2n}\\
\end{array}\right)
\end{equation}

The powers of the matrix $A$ can be found as follows:
Using the identity $\cos\arctan{x}=\frac{1}{\sqrt{1+x^2}}$
and defining the parameter $\omega=\arctan(2ws/c)$ we find
\begin{equation}
A=\left(
\begin{array}{cc}
c & 2sw \\
-2sw & c
\end{array}
\right)=\sqrt{c^2+4w^2s^2}
\left(
\begin{array}{cc}
\cos\omega & \sin\omega \\
-\sin\omega & \cos\omega
\end{array}
\right)
\end{equation}
and its powers are equal to
\begin{equation}
A^n=(c^2+4s^2w^2)^{n/2}\left(\begin{array}{cc}
\cos(n\omega) & \sin(n\omega)\\
-\sin(n\omega) & \cos(n\omega)
\end{array}
\right)\;.
\end{equation}

\begin{figure}
\begin{center}
\includegraphics[width=4cm]{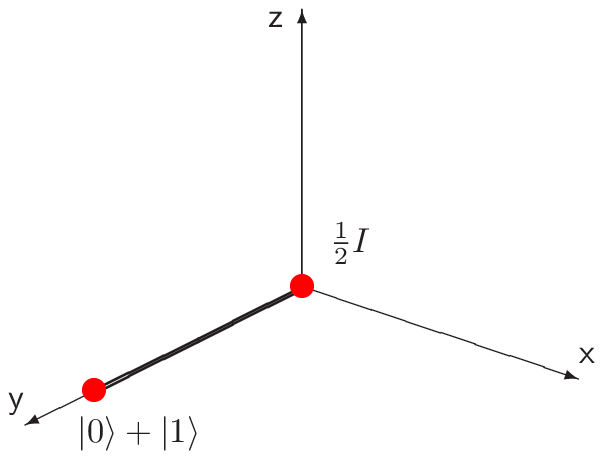}
\includegraphics[width=6cm]{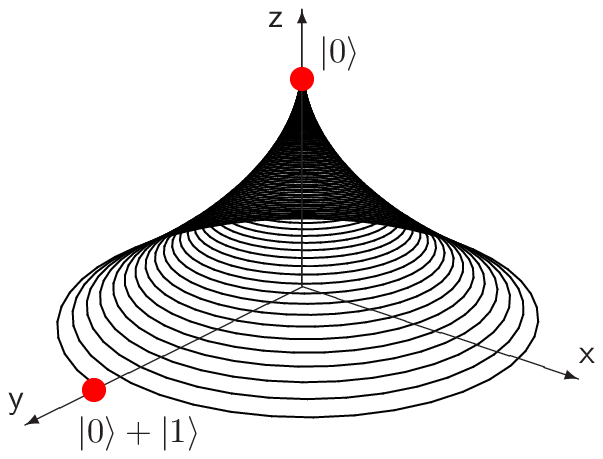}
\caption{Visualization of the evolution of a system qubit due to its interaction with reservoir
qubit in the process of quantum homogenization.
The system is supposed to be initially prepared in a pure state
$(|0\rangle+|1\rangle)/\sqrt{2}$.  The qubits in reservoir are in prepared in
the total mixture characterized by the values $w=0$ (left figure) and
and in the ground state $|0\rangle$ characterized by $w=1/2$ (right figure).
}
\label{homogenizacia}
\end{center}
\end{figure}

The introduction of the continuous time is now straightforward.
One has to replace $n$ with $t/\tau$. The dynamics
of open systems is usually characterized by two  parameters: the
{\it decay rate} $\Gamma_1$ and the {\it decoherence rate} $\Gamma_2$.
In our case we can introduce these parameters as follows
\begin{equation}
\begin{array}{lcl}
c^{2t/\tau}=e^{-\Gamma_1 t} & \Rightarrow
& \Gamma_1=1/T_1=-\frac{2}{\tau}\ln c\; ;\\
\left[c(c^2+4s^2 w^2)^{1/2}\right]^{t/\tau}=e^{-\Gamma_2 t} & \Rightarrow &
\Gamma_2=1/T_2=-\frac{1}{\tau}[\ln c\sqrt{c^2+4s^2w^2}]\; ,
\end{array}
\end{equation}
and  the continuous version of the homogenization process
can be described by a one-parametric set of maps
\begin{equation}
{\cal E}_t=\left(
\begin{array}{cccc}
1 & 0 & 0 & 0 \\
0 & e^{-\Gamma_2 t}\cos\Omega t & e^{-\Gamma_2 t}\sin\Omega t & 0 \\
0 & -e^{-\Gamma_2 t}\sin\Omega t & e^{-\Gamma_2 t}\cos\Omega t & 0 \\
2w(1-e^{-\Gamma_1 t}) & 0 & 0 & e^{-\Gamma_1 t}
\end{array}
\right)
\end{equation}
with frequency $\Omega=\omega/\tau$ describing the
``rotating'' part of the evolution (see Fig.\ref{homogenizacia}).
The general rule is that
before deriving  of the master equation itself one has to check that
the maps are valid quantum maps and that they fulfills the semigroup property.
However, here we will directly  derive the generator, and from its
properties the character of transformation will become obvious.
Using the method
described in  previous sections  we find out that the derived
generator is time independent
and it takes the form
\begin{equation}
{\cal G}=\left(
\begin{array}{cccc}
0 & 0 & 0 & 0 \\
0 & -\Gamma_2 & -\Omega & 0 \\
0 & -\Omega & -\Gamma_2 & 0 \\
2w\Gamma_1 & 0 & 0 & -\Gamma_1
\end{array}
\right)\, .
\end{equation}
Consequently, the master equation is of the Lindblad form with
{\it time-independent} coefficients $c_{jk}$ and $h_j$.
In particular, the non-vanishing parameters do read
$h_3=\Omega/2$, $c_{11}=c_{22}=\Gamma_1/4$, $c_{33}=(2\Gamma_2-\Gamma_1)/4$,
and $c_{12}=-i2w\Gamma_ 1$ and the matrix of $c_{jk}$
is positive whenever $\Gamma_2\ge\frac{1}{2}\Gamma_1$,
and $|w|\le 1/2$. The latter condition is trivially
satisfied, because of the positivity of $\xi$, i.e. $|w|\le 1/2$.
The first condition is related to a more general result (for details see Ref.~\cite{wodkiewicz}),
according to which
the decay is maximally two times faster than the decoherence,
i.e. $\Gamma_1\le 2\Gamma_2$, or equivalently $T_2\le 2T_1$. Its
validity in the case of homogenization can be checked by direct
calculation.

The positivity of the matrix $c_{jk}$ guarantees
the complete positivity of the whole evolution. The master equation
of the homogenization process reads
\begin{equation}
\begin{array}{l}
\dot{\varrho}_t=-i\frac{\Omega}{2}[S_3,\varrho_t]
+\frac{\Gamma_1}{4}(S_1\varrho S_1+S_2\varrho S_2-2\varrho)
+\frac{2\Gamma_2-\Gamma_1}{4}(S_3\varrho S_3-\varrho)\\
\hspace*{2cm}-iw\Gamma_1(S_1\varrho S_2-S_2\varrho S_1
+i\varrho S_3+iS_3\varrho)\; ,
\end{array}
\end{equation}
where the operators $S$ are given by Eq.(10).
Let us note that the parameter $\Omega$ describes
only the Hamiltonian part of evolution, which is in accordance with
our expectations. The Hamiltonian part causes the rotation, whereas
the other part of the generator induces a contraction into the
fixed point $\xi$. Only in the case when the homogenization
is completed the map cannot be inverted, but strictly speaking
this happens in the limit of time goes to infinity.

Let us make the following choice of parameters: $w=-1/2$,
$\Gamma_1=2\Gamma_2=2\gamma$. In this case the master equation
takes a simple form
\begin{equation}
\dot{\varrho}_t=i\frac{\Omega}{2}[S_3,\varrho]+
\gamma[2 S_- \varrho S_+ - S_- S_+ \varrho - \varrho S_-S_+]\; ,
\end{equation}
where $S_\pm=\frac{1}{2}(S_1\pm S_2)$. This is the well known
equation  describing the process of exponential decay
with the Hamiltonian $H=-\frac{\Omega}{2}S_3$ \cite{Nielsen2000}.

\section{Case study II.: Decoherence from collisions}
In this section we will analyze a collision-like dynamics which
models a decoherence of a qubit.
A more general treatment of decoherence
will be presented elsewhere \cite{ziman04}, while here
we will concentrate to the derivation of the master equation.
The task will be the same as before,
except that instead  of the partial-swap operation we will consider a
{\it partial CNOT} (controlled-NOT operation), i.e. $U_\eta=\cos\eta{\it I}+i\sin\eta\,{\tt CNOT}$.
The ${\tt CNOT}$ gate performs the $\sigma_x$ rotation
on the {\it target qubit}, when the {\it control qubit}
is in the state $|1\rangle$. If the control is in the state
$|0\rangle$, then the state of the target qubit is not changed.
Unlike the swap operation, the controlled NOT is asymmetric under the exchange
of qubits. Therefore, we have two different evolutions determined by
the role of the system qubit: it can be either the target qubit,
or the control qubit.

Let us assume the system qubit acts as the target. In this case the interaction
induces the map on the target qubit that reads
\begin{equation}
{\cal E}_\xi=\left(
\begin{array}{cccc}
1 & 0 & 0 & 0 \\
0 & 1 & 0 & 0 \\
0 & 0 & 1-2s^2\xi_{11} & 2cs\xi_{11} \\
0 & 0 & -2cs\xi_{11} & 1-2s^2\xi_{11} \\
\end{array}
\right)\; ,
\end{equation}
where we used again the notation $c=\cos\eta,s=\sin\eta$
and $\xi_{11}$ stands for $\langle 1|\xi|1\rangle$, i.e.
it is the parameter that characterizes the initial state of  reservoir qubits.
For the calculation of the powers of this map we will use
the same approach as before, because again we need to calculate the powers
of 2x2 matrix $A$ of the specific form
\begin{equation}
\label{A}
A=\left(
\begin{array}{cc}
a & b \\
-b & a
\end{array}
\right)=\sqrt{a^2+b^2}\left(
\begin{array}{cc}
\cos\omega & \sin\omega \\
-\sin\omega & \cos\omega
\end{array}
\right)=\sqrt{a^2+b^2} X(\omega)\; .
\end{equation}
In the last equality we exploited the identity
$\cos\arctan x=\frac{1}{\sqrt{1+x^2}}$  and we set $\omega=\arctan(b/a)$
to obtain the form suitable for evaluating the powers. Note that
$X(\omega)^n=X(n\omega)$. In our particular case
$a=1-2s^2\xi_{11}$ and $b=2cs\xi_{11}$. That is,
\begin{equation}
{\cal E}_\xi^n=\left(
\begin{array}{ccc}
1 & 0 & 0 \hspace{1cm} 0 \\
0 & 1 & 0 \hspace{1cm} 0 \\
\begin{array}{c}0\\ 0\end{array}
& \begin{array}{c}0\\ 0\end{array}  &
(1+4s^2\xi_{11}\xi_{00})^{n/2} X(n\omega)
\end{array}
\right)\; .
\end{equation}
Following the same methods as in the previous section we can derive the generator of the process ${\cal G}$ in
a very simple form
\begin{equation}
{\cal G}=\left(
\begin{array}{cccc}
0 & 0 & 0 & 0 \\
0 & 0 & 0 & 0 \\
0 & 0 & -\Gamma & \Omega \\
0 & 0 & -\Omega & -\Gamma
\end{array}\right) \Rightarrow
\dot{\varrho}_t=-i\frac{\Omega}{2}[\sigma_x,\varrho_t]
+\frac{\Gamma}{2}(\sigma_x\varrho_t\sigma_x-\varrho_t)\; .
\end{equation}
As seen from its structure this generator describes
a ``pure'' decoherence, i.e. a process of diagonalization
of the state in the basis associated with the eigenvectors of the operator
$\sigma_x$. The parameters of the dynamics are defined in a similar
way as before, i.e.
$\Omega=\omega/\tau=\arctan\frac{2cs\xi_{11}}{1-2s^2\xi_{11}}$ and
$\Gamma=-\frac{1}{\tau}\ln\sqrt{1-4s^2\xi_{11}\xi_{00}}$.

If we consider the system qubit to play the role of the control, then the
dynamical map ${\cal E}_\xi$ is very similar to the previous case, i.e.
\begin{equation}
{\cal E}=\left(
\begin{array}{cccc}
1 & 0 & 0 & 0 \\
0 & c^2+s^2\langle\sigma_x\rangle_\xi &
cs(1-\langle\sigma_x\rangle_\xi) & 0 \\
0 & -cs(1-\langle\sigma_x\rangle_\xi) &
c^2+s^2\langle\sigma_x\rangle_\xi & 0 \\
0 & 0 & 0 & 1 \\
\end{array}
\right)\; ,
\end{equation}
where $\langle\sigma_x\rangle_\xi={\rm Tr}\xi\sigma_x$. Defining
the parameters $\Gamma=
-\frac{1}{\tau}\ln\sqrt{c^2+s^2\langle\sigma_x\rangle_\xi}$,
$\Omega=\frac{1}{\tau}\arctan{\frac{cs(1-\langle\sigma_x\rangle_\xi)}
{c^2+s^2\langle\sigma_x\rangle}}$ we can directly write down the
master equation
\begin{equation}
\dot{\varrho}_t=-i\frac{\Omega}{2}[\sigma_z,\varrho_t]
+\frac{\Gamma}{2}(\sigma_z\varrho_t\sigma_z-\varrho_t)\; .
\end{equation}
Likewise in the previous case we obtain the dynamics describing the decoherence.
Nevertheless, there is a difference, if the system qubit plays the role of the control, then
the decoherence is observed in
the basis associated with the eigenvectors
of the $\sigma_z$ operator, rather then in the basis of eigenvectors $\sigma_x$ that was the case when the
system qubit plays the role of the target. We see that irrespective on the fact whether the system qubit
is a controle or a target the sequence of partial CNOT collisions leads to decoherence processes. There are two differences though.
The first one is the basis in which the decoherence takes place and the second is the decoherence time. In particular, the decoherence rates depend on
the initial state of the reservoir $\xi$ in different ways.

\section{Conclusion}
We have shown that using a simple collision-like model one can derive master equations
for a qubit interacting with an environment. In our approach
the discrete dynamics is described by a
sequence of unitary transformations representing bi-partite interactions (``collisions'').
As a result of the collision-like evolution the induced one-qubit dynamics
is discrete, too. However, we have shown that
in specific cases (the partial SWAP operation and the partial CNOT operation)
this essentially discrete evolution can be substituted by a
continuous dynamics. The resulting time evolution fulfills the semigroup property.
Using the state-space description of the dynamics of a qubit we can say that
the evolution of a qubit state that undergoes  a sequence of collisions
can be illustrated as a ordered sequence of points in the
Bloch sphere. We have shown how to connect these points with
a smooth line representing the continuous time evolution driven by
a Lindblad master equation.

The presented approach can be generalized
to higher-dimensional quantum systems (qudits), as well as to collisions
described by more general bi-partite unitary transformations $U$. The method
essentially works for any interaction which induce an
{\it invertible} map $\cal E$ (i.e. $\det {\cal E}\ne 0$).
It turns out that semigroups ${\cal E}_t=e^{-{\cal G}t}$
always contain only invertible maps, because for the (usually unphysical)
map ${\cal E}_{-t}$ the following identity holds
${\cal E}_t{\cal E}_{-t}=e^{-{\cal G}t}e^{+{\cal G}t}=e^{0}={\cal I}$.

Another (open) problem concerning the derivation of the master equation from a discrete collision-like dynamics
is whether for all invertible mappings $\cal E$, the semigroup
generated by its powers can be always interpolated with a continuous
semigroup of completely positive maps. This problem leads to
the question about the structure of the set of quantum semigroups.
It is worth to study which bi-partite interactions in the
collision-like models could
stand behind the known quantum processes described
approximatively by quantum master equations. We believe that this approach
of derivation of the master equations
provides us with a new insight into the dynamics of open systems.

\noindent
{\bf Acknowledgements}\newline
This was work supported in part by  the European
Union projects QUPRODIS  and CONQUEST.


\begin{thebibliography}{10}
\bibitem{lindblad76}
G. Lindblad,
{\it On the generators of Quantum Dynamical Semigroups},
{\it Commun. Math. Ph.} {\bf 48}, 119-130 (1976)

\bibitem{davies}
E.B.Davies, {\it Quantum Theory of Open Systems},
(Academic, London, 1976)

\bibitem{alicki}
R.Alicki, K.Lendi, {\it Quantum Dynamical Semigroups and Applications},
Lecture notes in Physics (Springer-Verlag, Berlin, 1987)

\bibitem{Nielsen2000}
M.A.Nielsen and I.L.Chuang
{\it Quantum Computation and Quantum information}
(Cambridge University Press, Cambridge,

\bibitem{spohn}
H.Spohn, {\it Kinetic equations from Hamiltonian dynamics: Markovian limit},
{\it Rev.Mod.Phys.} {\bf 53}, No.3, 569 -- 615 (1980)

\bibitem{Ziman012}
M.Ziman, P.\v Stelmachovi\v c, V.Bu\v zek, M.Hillery, V.Scarani, and N.Gisin,
{\it Diluting quantum information: An analysis of information transfer in system-reservoir interactions},
{\it Phys.Rev A} {\bf 65}, 042105 (2002),
LANL preprint archive {\tt quant-ph/0110164}

\bibitem{Scarani01}
V.Scarani, M.Ziman, P.\v Stelmachovi\v c, N.Gisin, and V.Bu\v zek,
{\it Thermalizing quantum machines: Dissipation and Entanglement},
{\it Phys. Rev. Lett.} {\bf 88}, 097905 (2002),
LANL preprint archive {\tt quant-ph/0110088}

\bibitem{Ziman2003}
M.Ziman, P.\v Stelmachovi\v c, and V.Bu\v zek,
{\it Saturation of Coffman-Kundu-Wootters inequality via quantum homogenization.}
{\em J. Opt. B: Quantum Semiclass} {\bf 5}, s439  (2003).


\bibitem{Buzek1996}
V.\ Bu\v{z}ek and M.\ Hillery,
{\it Quantum copying: Beyond the No-Cloning Theorem}
{\it Phys.\ Rev.\ A} {\bf 54}, 1844 (1996).

\bibitem{wodkiewicz}
S.Daffer, K.Wodkiewicz, J.K.McIver,
{\it Bloch equations and completely positive maps},
{\it J. Mod. Opt.} {\bf 51 }, 1843 (2004),
LANL preprint archive {\tt quant-ph/04011177}


\bibitem{ziman04}
M.Ziman et al., in preparation

\end{thebibliography}
\end{document}